# Accessibility Considerations in the Development of an AI Action Plan

This is a response to the OSTP Request for Information (RFI) on the Development of an Artificial Intelligence (AI] Action Plan was published in the Federal Register: https://www.federalregister.gov/documents/2025/02/06/2025-02305/request-for-information-on-the-development-of-an-artificial-intelligence-ai-action-plan.

## Key Positions

We argue that there is a need for Accessibility to be represented in several important domains:
- Capitalize on the new capabilities AI provides
- Support for open source development of AI, which can allow disabled and disability focused professionals to contribute, including
  - Development of Accessibility Apps which help realise the promise of AI in accessibility domains
  - Open Source Model Development and Validation to ensure that accessibility concerns are addressed in these algorithms
  - Data Augmentation to include accessibility in data sets used to train models
  - Accessible Interfaces that allow disabled people to use any AI app, and to validate its outputs
  - Dedicated Functionality and Libraries that can make it easy to integrate AI support into a variety of settings and apps.
- Data security and privacy and privacy risks including data collected by AI based accessibility technologies; and the possibility of disability disclosure.
- Disability-specific AI risks and biases including both direct bias (during AI use by the disabled person) and indirect bias (when AI is used by someone else on data relating to a disabled person).

## 1. Introduction

AI has the potential to *empower everyone* to become more independent and self-sufficient. The increasing use of artificial intelligence (AI)-based technologies in everyday settings creates new opportunities to understand how disabled people might use these technologies [Glazko, 2023]. It also enables the development of new types of assistive technologies as well as new ways for people with disabilities to interact with technology in ways that are both simpler (for those who need things simpler) and more efficient and effective for those who cannot use the traditional interfaces effectively. AI has been rapidly taken up in almost all accessibility communities [Adnin 2024, Alharbi 2024, Jiang 2024, Bennett 2024, Valencia 2023]. Since becoming widely available to the public, Generative Artificial Intelligence (GAI) has steadily gained recognition for its potential as a valuable tool in the private sector and by government, as well as a tool for accessibility. Studies of blind and visually impaired individuals have found that they use GAI to 'offload' cognitively demanding tasks and obtain personal help such as fashion advice (*e.g.,* [Xie 2024]), and to create content or retrieve information [Adnin 2024]. A study of GAI use by neurodiverse users found GAI can both support and complicate tasks like code-switching, emotional regulation, and accessing information [Glazko, 2025]. A study of people who use AAC found it helpful for text input [Valencia 2023].

However there are concerns with a technology that is often based on probability and thus tends toward the most common case rather than those at the margins. Recent reports by Whittaker et al. [2019], Trewin et al. [2019], and Guo et al. [2020] highlight concerns about AI's potential negative impact on inclusion, representation, and equity for those in marginalized communities, including disabled people. For example, if an autonomous vehicle fails to detect an unusual case, such as a wheelchair user who propels herself backwards using her feet [Moura 2022], that error could lead to a life or death situation. An AI might also associate disability with toxic content or harmful stereotypes [Mack 2024] or rate a resume lower because of presumed incompetence [Glazko, 2024]. These results may be due to lack of representation during data collection, or algorithms that learn primarily from common cases. These problems replicate and amplify biases experienced by disabled people when interacting in everyday life. Further, AI may fail to accessibly support verification and validation [Glazko, 2023].

Moving forward, our challenge will be to capitalize on the new capabilities AI provides, while avoiding the potential problems it might create due to biases and other concerns. In this RFI response, we define disability in terms of the discriminatory and often systemic problems with available infrastructure's ability to meet the needs of all people. We then highlight where disability opportunities and risks in critical arenas raised in the RFI in three areas: Open source development of AI apps, algorithms, and data sets that address accessibility; data security and privacy and privacy risks relating to accessibility; and disability-specific AI risks and biases including both direct bias (during AI use by the disabled person) and indirect bias (when AI is used by someone else on data relating to a disabled person).

One of the biggest benefits of the current generation of AI, generative models, is their ability to provide agency and control to people with disabilities. This concept of agency and control has been highlighted as critical to an accessible AI future that must be explicitly cultivated and supported, *"given the increasingly proprietary nature of the technologies being created, and the centralization inherent in the current form of AI"* [Whittaker, 2019]. To create this future, we must recognize and learn how to address the risks and biases that disabled people face.

## 2. Open Source Development of AI

Open source development is critical to reducing economic barriers and ensuring broad access. Further, it provides people with disabilities the opportunity to participate in the development of the technologies they use. They best understand their needs and many express interest in tailoring technologies to those specific needs - but they need access to the right tools to do so [Glazko 2023]. Open Source AI development can take several forms:

- **Accessibility App Development:** By creating open source opportunities for AI app development, people with disabilities who are interested have an avenue to be involved in the development process of their own assistive technologies. We are already seeing some of this happening in AI "app stores" such as the ChatGPT AI ecosystem, which allows the creation of 0-code GPTs, enabling non-programmers to create custom GPTs. In that ecosystem, numerous apps have been created for accessibility reasons. However, these apps are unverified and not necessarily easy to find.
- **Model Development and Validation for Accessibility:** If AI models are open sourced, it becomes possible to examine how models account for small group biases and other factors that might impact model reliability. Disabled people can contribute to innovation and extension of those models. Further, accessibility related validation is not well supported in AI models. Open sourcing model validation algorithms could allow developers to build on them by adding and attending to accessibility.
- **Data Augmentation to Include Accessibility:** If AI training data is open sourced, it becomes feasible for people with disabilities to contribute to that data and for everyone to examine that data to ensure that it represents them. This is critical to helping to reduce model bias
- **Accessible Interfaces:** Relatedly, any model that is being used by the public must have some sort of user interface. Open source interfaces can be made accessible, something that is not necessarily supported in enterprise models. Interface design must also support accessible validation, such as the ability of a disabled person to assess whether or not the source and the result of their query are aligned when one or both of those were not accessible in the first place.
- **Dedicated Functionality and Libraries:** Special functions that could extend the capabilities of assistive technology for a wide range of disabilities could be developed using open-source code and open-AI. This would allow a widespread shift of what is possible across a wide range of assistive technologies. For example, power wheelchair developers could add speech commands with minimal effort.

# 3. Data Privacy and Security implications

Data privacy and security can influence how AI might disclose or impact disability disclosure. Disability status is increasingly easy to detect from readily available data such as mouse movements [Youngmann 2019]. Any system that can detect disability can also track its progression over time, possibly before a person knows they have a diagnosis. This information could be used, without consent or validation, to deny access to housing, jobs, or education, potentially without the knowledge of the impacted individuals [Whittaker, 2019]. Additionally, AI biases may require people with specific impairments to accept reduced digital security, such as the person who must ask a stranger to "forge" a signature at the grocery store " … because I can't reach [the tablet]. [Kane 2020]" This is not only inaccessible, it is illegal: kiosks and other technologies such as point-of-sale terminals used in public accommodations are covered under Title III of the Americans with Disabilities Act.

Privacy and security are particularly important when people with disabilities need to rely on technology or rely on technology for a wide array of functions. For example, individuals who cannot rely on speech to be heard and understood often use AAC technology to communicate effectively across a wide range of life contexts. Similarly, people who use captions often rely on AI to help them access speech in conversations with others. Unlike others, individuals with communication-related disabilities may not have the option of communicating highly sensitive information without making use of technology. For this reason, it is critical that data privacy be emphasized, and that individuals with disabilities have options around when or how their communication is being stored/shared [Williams 2024, Sellwood 2024]. For example, human operators providing Relay calls for the deaf are currently bound by strict confidentiality requirements, allowing users to have confidence that their credit card information and medical histories will not be shared.

The shift from cloud based to locally run AI holds the promise for allowing AI to be used in ways that cannot leak information back to the cloud. This can allow people with disabilities to tap the power of AI for personal agents or for use in assistive technologies without the privacy risks that accompany the use of cloud based AI. This requires general advances in AI but also specific research and development to create versions of locally operated AI that will provide the specific AT features needed. One strategy that has been discussed is to use cloud based AI for now and switch to local AI when it is possible. It should be noted however that once a person's information is in the cloud, moving to local AI will not remove the information already released. Therefore, the move to local AI is an important imperative for those whose personal information could be used in ways not beneficial to them.

---

## *Callout: Communication Technologies*

Communication technologies should provide transparent information to people with disabilities about data use:
- What of my data is being used?
- By whom?
- What is my data being used for?

This information must be presented in a clear concise manner to be easily understood. Such technology should also provide people with autonomy / control over the use of their data. These controls must be easily understood and readily apparent. Specifically, they ask that it be easier to opt in and out using the following methods:
- Cleary display messages with information on data access and usage and then ask the user if they want to opt-out.
- Display the message anytime data access or usage changes.
- Make it simpler to temporarily turn off data collection.

Even if these technologies begin to move onto local devices, the need for security and transparency may still persist. For example, when one person in a zoom call uses transcription, everyone in the call is transcribed.

# 4. Disability-Specific AI Risks and Biases

Even the most well-designed of systems may cause a variety of harms when deployed. It is critical that technologists learn about these harms and how to address them before deploying AI-based systems. AI model development must be extended to consider risks to disabled people. Past works have highlighted harms caused by problems such as unrepresentative data [Whittaker, 2019], missing and unlabeled data [Glazko 2024, Guo 2020], measurement errors [Trewin 2019], and AI that exacerbates or causes disability [Whittaker, 2019], or limits access to care, thus exacerbating a disability [Lecher 2018]. Further, even if an AI-based system is carefully designed to minimize bias, the interface to that algorithm, its configuration, the explanation of how it works, or the potential to verify its outputs may be inaccessible.

As noted above, once a person's information – particularly information about their disabilities – has been leaked to the cloud due to unanticipated leaks, it is not sufficient to plug the leaks going forward. Once the information is out, it is out for good, or rather for good or bad. Hence much care needs to be taken from day 1 in the use of AI.

## 4.1 Incorrect, misleading inferences and hallucinations.

Even when AI models are trained on representative data, inferences based on these models are prone to mistakes, misleading interpretations of the data and outright hallucinations. We address each of these in turn.

- **Errors:** An AI may make a mistake, such as confusing a door for a book in an indoor scene or providing incorrect advice about a medical or legal situation. Similarly, when simplifying text, a GAI might introduce discrepancies.
- **Misleading Interpretations and Misinformation:** An AI may be "correct" but still misleading. For example, imagine a scenario in which optical character recognition (OCR) on a photo has perfect recognition of the text visible in the photo, but lacks important context that drastically changes how this text should be interpreted.
- **Erasure:** An AI may help with a communication task, supporting an author while simultaneously erasing some of their identity such as cultural or language expression even as it supports clarity or ease of communication. Or it may only support some aspects of a person's identity, such as captioning that supports only one language at a time, or makes assumptions about language fluency [Desai 2025].
- **Hallucination:** GAI can fabricate false information due to inaccurate, incomplete, or biased training data. For example, GAI tools try to predict what words should come next in a conversation you are having with them. This can lead to sentences that sound correct, but were formed without an understanding of the meaning behind the words. An example is hallucinating a bus stop in a photo of an outdoor scene that lacks a bus stop.

Appropriate usage of AI for any purpose should include consideration of these risks and practical, *accessible* ways of verifying the AI inferences whenever possible. When either the input to the AI, or the output from the AI, are inaccessible, this can make these risks even larger because verification is inaccessible. User training is also needed, e.g., to deal with AI behavior such as confident-sounding answers to user queries even when the answers are wrong [AFB 2025].

The use of AI that is restricted in its answers to a set of reliable data can guard against this. This is a strategy being used more and more to achieve more reliable answers.  There are limits to this however and it does not allow for questions on all topics if its answers are restricted to only a curated set of data.

## 4.2 AI bias against people with disabilities

AI bias against people with disabilities has been documented in AI systems, and can occur both in direct use [Mack 2024] and when used by someone other than the disabled person themselves [Glazko 2024].

**Direct Bias:** AI may associate disability with toxic content or harmful stereotypes [Mack 2024]. For example, AI may depict disabled people as lonely, sad, old, or even horrific (see image).

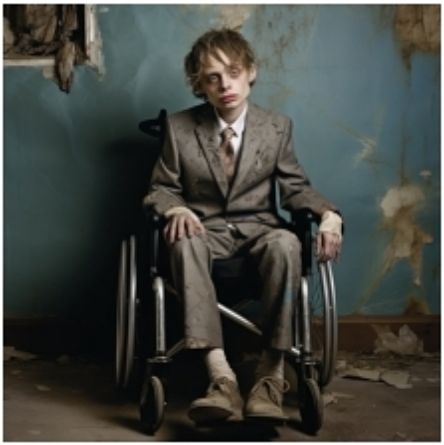

(Caption: A picture of a person in a wheelchair with blank eyes, dirty clothing, a gaunt face, and a peeling wall behind him. Image taken from [Mack 2024])

**Indirect Bias:** As an example of indirect bias, consider resume screening, which is not done by a jobseeker directly, but rather by the company they apply to. When that company uses AI, this is indirectly affecting the job seeker. A study of ChatGPT bias compared a resume that mentions disability in a leadership award, scholarship and presentation in comparison to the same resume with those items removed. ChatGPT usually picked the resume without the prestige items, and the degree of bias varied with which disability was mentioned [Glazko, 2024].

As stated earlier, these biases can have a variety of causes, including biases in who is included during data collection (such as lacking speech data that represents people with dysarthria), biases in the data itself (such as biased statements about people with disabilities), and algorithmic biases (such as privileging common cases over outliers). Thus addressing them will require a multifaceted approach that addresses biases in the fundamental principles and mechanisms on which GAI is based (as discussed in Section 1, open source development) as well as regulation and governance (discussed next).

# 5. Regulation and Governance

Regulation and governance are necessary to support peoples rights in fighting AI bias and to reduce bias. This is especially important because of the multitude of ways in which AI can determine how disability governance is enacted, as discussed in past work [Kane 2020, Whittaker 2019]. For example, AI may only recognize some bodies as human [Guo, 2020; Kane 2020], or as disabled [Whittaker, 2019]. However, the Americans with Disabilities Act does not require a diagnosis for someone to count as disabled, instead according that label even to people who are simply treated as disabled, or have a history that would be recognized as disabled without their assistive technology (42 U.S.C. § 12101 (a)(1)). How can AI technologies detect these nuances? The risk of misidentification includes serious consequences such as denying access to services.

Mankoff et al [2024] argue for setting standards regarding whether and how algorithms are assessed for accessibility and for errors relating to accessibility. Relatedly, they argue, *"consumer consent and oversight concerning best practices are both essential to fair use. AI-based systems should be interpretable, overridable, and support accessible verifiability."* This will require an investment in research and oversight to ensure compliance. For example, we must develop AI benchmarks that allow the developers of AI applications to test whether they are sufficiently unbiased to be safe for deployment. These benchmarks can minimize the deployment of biased AI. Relatedly, research is needed to identify ways that the inherent bias of AI toward the common case can be avoided through improved algorithms.

# 6. Conclusions and Recommendations

*"First and foremost, do no harm: algorithms that put a subset of the population at risk should not be deployed."* [Mankoff 2024].

To accomplish this goal, we need to make the most of the positive future AI can support, while avoiding its most deleterious effects. This can occur through a combination of regulation, research, and innovation. However the least effective of these is likely to be regulation due to the rapid pace of advance and the slow page of legislation. Hence the need for research and innovation to complement regulation as our primary tools for addressing these issues.

This in turn will call for a change in who has access to become builders of AI, including access to higher education, leadership and so on [Tadimalla 2024, Mankoff 2022]. We need to ensure everyone is included in all of these domains, and included in the target audiences for AI. This includes groups and individuals who are disabled (e.g., people who need or use AAC, people with intellectual disabilities), from varied backgrounds, varied socioeconomic status, *etc*. Further, it is essential that we continue to monitor AI for new and evolving problems and challenges and develop strategies to address them before new systems come online.

There is nothing inherent to technology generally (and AI specifically) that makes it inaccessible. Rather, it is due to our design of the technology and the care that we take, or do not take, that problems with its accessibility and its fairness may occur.

# Bibliography

[Adnin 2024] Adnin, Rudaiba, and Maitraye Das. (2024) "'I look at it as the king of knowledge': How Blind People Use and Understand Generative AI Tools." Proceedings of the 26th International ACM SIGACCESS Conference on Computers and Accessibility. 2024.

[AFB 2025] "Empowering or Excluding: New Research and Principles for Inclusive AI", American Foundation for the Blind. Jan 2025. https://afb.org/research-and-initiatives/empowering-or-excluding/narrative-and-findings

[Alharbi 2024] Alharbi, R., Lor, P., Herskovitz, J., Schoenebeck, S., & Brewer, R. (2024). Misfitting With AI: How Blind People Verify and Contest AI Errors. arXiv preprint arXiv:2408.06546.

[Bennett 2024] Bennett, C. L., Shelby, R., Rostamzadeh, N., & Kane, S. K. (2024, October). Painting with Cameras and Drawing with Text: AI Use in Accessible Creativity. In The 26th International ACM SIGACCESS Conference on Computers and Accessibility (pp. 1-19).

[Desai 2025] Desai, A. et al. (2025) Toward Language Justice: Exploring Multilingual Captioning for Accessibility. CHI 2025. https://doi.org/10.1145/3706598.3713622

[Glazko 2023] Glazko, K.S. et al. (2023) An autoethnographic case study of generative artificial intelligence's utility for accessibility. In Proceedings of the 25th Intern. ACM SIGACCESS Conf. on Computers and Accessibility, ASSETS 2023, New York, NY, USA, (October 22–25, 2023); https://bit.ly/484udWp

[Glazko 2024] Glazko, K.S. et al. (2024) Identifying and improving disability bias in GAI-based resume screening. In FaCCT 2024; https://bit.ly/489R5nd.

[Glazko 2025] Glazko, K. S. et al. (2025) Autoethnographic Insights from Neurodivergent GAI "Power Users". CHI 2025. https://doi.org/10.1145/3706598.3713670

[Guo 2020] Guo, A. et al. (2020) Toward fairness in AI for people with disabilities SBG@a research roadmap. ACM SIGACCESS Access. Comput. 125, 2; https://bit.ly/3U5cGrl

[Jiang 2024] Jiang, L., Jung, C., Phutane, M., Stangl, A., & Azenkot, S. (2024, May). "It's Kind of Context Dependent": Understanding Blind and Low Vision People's Video Accessibility Preferences Across Viewing Scenarios. In Proceedings of the CHI Conference on Human Factors in Computing Systems (pp. 1-20).

[Kane 2020] Kane, S.K. et al. (2020) Sense and accessibility: Understanding people with physical disabilities' experiences with sensing systems. In ASSETS '20: The 22nd Intern. ACM SIGACCESS Conf. on Computers and Accessibility. T.J. Guerreiro, H. Nicolau, and K. Moffatt, Eds. Virtual Event, Greece, (Oct. 26–28, 2020); https://bit.ly/3UbgK9l

[Lecher 2018] Lecher, C. (2018) What happens when an algorithm cuts your health care. The Verge 21, 3.

[Mack 2024] Mack, Kelly Avery, et al. (2024) "'They only care to show us the wheelchair': Disability representation in text-to-image AI models." Proceedings of the 2024 CHI Conference on Human Factors in Computing Systems.

[Mankoff 2022] J. Mankoff, D. Kasnitz, D. Studies, L. J. Camp, J. Lazar, and H. Hochheiser. Areas of Strategic Visibility: Disability Bias in Biometrics. Tech. rep. Office of Science, Technology Policy Notice of Request for Information (RFI) on Public, and Private Sector Uses of Biometric Technologies, 2022. arXiv: 2208.04712.

[Mankoff 2024] Mankoff, Jennifer, et al. "AI Must Be Anti-Ableist and Accessible." Communications of the ACM 67.12 (2024): 40-42.

[Moura 2022] Ian Moura, "Addressing Disability and Ableist Bias in Autonomous Vehicles: Ensuring Safety, Equity and Accessibility in Detection, Collision Algorithms and Data Collection" Disability Rights Education & Defense Fund https://dredf.org/addressing-disability-and-ableist-bias-in-autonomous-vehicles-ensuring-safety-equity-and-accessibility-in-detection-collision-algorithms-and-data-collection/

[Sellwood 2024] Sellwood, D., et al. (2024). Imagining alternative futures with augmentative and alternative communication: a manifesto. Medical Humanities, 50(4), 620-623.

[Tadimalla 2024] Tadimalla, Sri Yash, Rachel Figard, and Yukyeong Song. "WIP: Moving from Accessibility to Anti-Ableism through the Explication of Disability in the AI Ecosystem." 2024 IEEE Frontiers in Education Conference (FIE). IEEE, 2024.

[Trewin 2019] Trewin, S. et al. (2019) Considerations for AI fairness for people with disabilities. AI Matters 5, 3; https://bit.ly/4f5ur1A

[Whittaker, 2019] Whittaker, M. et al. (2019) Disability, bias, and AI. AI Now Institute 8.

[Xie, 2024] Xie, J. et al. (2024). Emerging practices for large multimodal model (LMM) assistance for people with visual impairments: Implications for design. arXiv preprint arXiv:2407.08882 (2024)

[Youngmann 2019] Youngmann, B. et al. (2019) A machine learning algorithm successfully screens for Parkinson's in web users. Annals of Clinical and Translational Neurology 6, 12.

[Williams 2024] Williams, K., & Holyfield, C. (2024, May). Future of AAC technologies: Priorities for inclusive research and implementation [Oral presentation]. The Future of AAC Research Summit, Arlington, VA, United States.

[Valencia, 2023] Valencia, S., Cave, R., Kallarackal, K., Seaver, K., Terry, M., & Kane, S. K. (2023, April). "The less I type, the better": How AI language models can enhance or impede communication for AAC users. In Proceedings of the 2023 CHI Conference on Human Factors in Computing Systems (pp. 1-14). https://doi.org/10.1145/3544548.3581560

# Signatories

Jennifer Mankoff, Director, the RERC on ICT and Accessibility (CREATE)
Janice Light, Christine Holyfield, & Erik Jakobs, The RERC on AAC
James Coughlan, the RERC on Blindness and Low Vision
Christian Vogler and Abraham Glasser, Co-Directors, the RERC on Deaf and Hard of Hearing Technology
Gregg Vanderheiden, TRACE RERC, University of Maryland, and Raising the Floor
Laura Rice, Director, Technologies to Support Aging Among Older Adults with Long-Term Disabilities (TechSAge)